\documentclass[a4paper]{article}

\usepackage{INTERSPEECH2022}
\usepackage{subfigure}
\usepackage{multirow}
\usepackage{float}
\usepackage{color}
\usepackage{stfloats}

\usepackage{enumitem}

\title{An Anchor-Free Detector for Continuous Speech Keyword Spotting}
\name{Zhiyuan Zhao$^1$, Chuanxin Tang$^1$, Chengdong Yao$^2$, Chong Luo$^1$}
\address{
  $^1$Microsoft Research Asia\\
  $^2$University of Technology Sydney}
\email{zhiyzh@microsoft.com, chutan@microsoft.com, chengdong.yao-1@student.uts.edu.au, cluo@microsoft.com}

\begin{document}

\maketitle
\begin{abstract}
    Continuous Speech Keyword Spotting (CSKWS) is a task to detect predefined keywords in a continuous speech. In this paper, we regard CSKWS as a one-dimensional object detection task and propose a novel anchor-free detector, named AF-KWS, to solve the problem. AF-KWS directly regresses the center locations and lengths of the keywords through a single-stage deep neural network. In particular, AF-KWS is tailored for this speech task as we introduce an auxiliary unknown class to exclude other words from non-speech or silent background. We have built two benchmark datasets named LibriTop-20 and continuous meeting analysis keywords (CMAK) dataset for CSKWS. Evaluations on these two datasets show that our proposed AF-KWS outperforms reference schemes by a large margin, and therefore provides a decent baseline for future research.  

\end{abstract}

\section{Introduction}

Keyword spotting is a classical problem in speech recognition. The most well-known application is trigger word detection for smart speakers and smart devices, such as ``Okay Google'' \cite{wake-google-dnn,wake-google-e2e} and ``Hey Siri'' \cite{wake-apple-1dconv,wake-apple-accVSlatency,wake-apple-ctc}. Another widely discussed application is speech command recognition for remote control, such as ``Yes'', ``No'', ``Up'', and ``Down'' \cite{data-speech-commands}. In this paper, however, we are interested in continuous speech keywords spotting (CSKWS), which finds many applications in prohibited words filtering and recorded meeting understanding. The latter is attracting increasing interests due to the explosive increase of online meetings and meeting recordings since the COVID-19 pandemic. 
By defining a small set of meeting-related keywords and key phrases, such as ``Start'', ``Agenda'', ``To-do'', and ``Summarize'' and detecting them with a dedicated model, one can divide a meeting into chapters in a cost-effective way, compared to running automatic speech recognition (ASR) or large-vocabulary continuous speech recognition (LVCSR) throughout the meeting recording. 

Existing technologies and datasets developed for trigger word detection\cite{ckws-wake-rpn} and speech command recognition \cite{trim-google-streaming,trim-matchbox} are not readily applicable to CSKWS. In trigger word detection, the objective is to detect a single pre-defined trigger word or trigger phrase on an edge device.
In speech command recognition, it is often assumed that the keywords are solitary, with clear and easily detectable boundaries with other words. 
In Google Speech Commands dataset \cite{data-speech-commands}, each one-second-long speech sample contains one separated keyword. 
The main challenge of CSKWS, which makes it a unique research problem, lies in the presence of multiple target keywords and the interference of context in continuous speech. 

In this paper, we attempt to standardize the definition, benchmark dataset, and evaluation criteria of the CSKWS problem. Previously, Rostami et al. \cite{ckws-football} collected the football keyword dataset (FKD) for CSKWS, but they only collected the lone word audio and then synthesized the continuous speech. As only keywords are authentic speech, one could expect a simple authentic/synthetic speech discriminator to perform well on FKD. 
We propose two benchmark datasets for the CSKWS problem. The first one, named LibriTop-20, is derived from the LibriSpeech dataset \cite{data-librispeech}. It contains 20 most frequently appeared words with at least two syllables. There are a total of 256k utterances and 530k keywords. The second one is a brand new dataset named continuous meeting analysis keywords (CMAK). We define 24 meeting-structure-related keywords, such as ``Agenda'', ``Start'', and ``Question''. We collect over 15 sentences for each keyword and employ 32 human voices and 136 synthetic voices to produce 350k utterances.  
 
We present a novel algorithm for the CSKWS problem. Specifically, we treat CSKWS as an object detection problem rather than a classification problem. 
{In computer vision (CV), anchor-based object detectors utilize predefined anchors to enumerate possible locations, scales, and aspect ratios which introduce redundant computing, while anchor-free object detectors directly learn the object existence possibility and bounding box coordinates. }
Inspired by the work in CV \cite{other-centernet,other-od-af,other-od-af2}, we propose an anchor-free method, named AF-KWS, for speech keyword spotting. 
We predict the center location of a keyword in continuous speech and regress the length of the keyword by a simple convolutional head. An additional offset head is employed to refine the predicted center position. While these designs are directly extended from visual object detection, we introduce an auxiliary ``unknown" class for the interfering words in continuous speech. Such a design differentiates interfering words from the non-word background containing silence and environment noise, and thus significantly improves the accuracy of the keyword spotting model. 

We evaluate the proposed AF-KWS method on the two benchmark datasets we have built. In addition to the commonly adopted evaluation metrics, such as false rejection rate (FRR), false alarms (FAs) per hour, and accuracy, we also adopt the object detection metric mAP. We compare AF-KWS with two reference methods adapted from the SOTA trigger word detection algorithms by an inserted sliding-window region proposal module. Results show that, AF-KWS outperforms these adapted methods by a large margin. This justifies our claim that CSKWS is a unique research problem and should not be viewed as a simple extension of trigger words detection or speech command recognition. We will release our benchmark datasets \footnote{Utterances of human voice will be excluded from CMAK to avoid privacy concerns.} as well as the codes of AF-KWS. We hope our work can pave the way for future research on this interesting CSKWS problem.

\section{Anchor-Free Keywords Spotting}

\subsection{Overview of AF-KWS}
CSKWS is treated as a 1D object detection problem. Given a predefined keywords set $K=\{k_1, k_2, ..., k_C\}$ of size $C$ and an input audio of length $r$, the task of CSKWS is to find the locations and lengths of all the keywords in the input audio. 

\begin{figure*}[t]
  \centering
  \includegraphics[width=\linewidth]{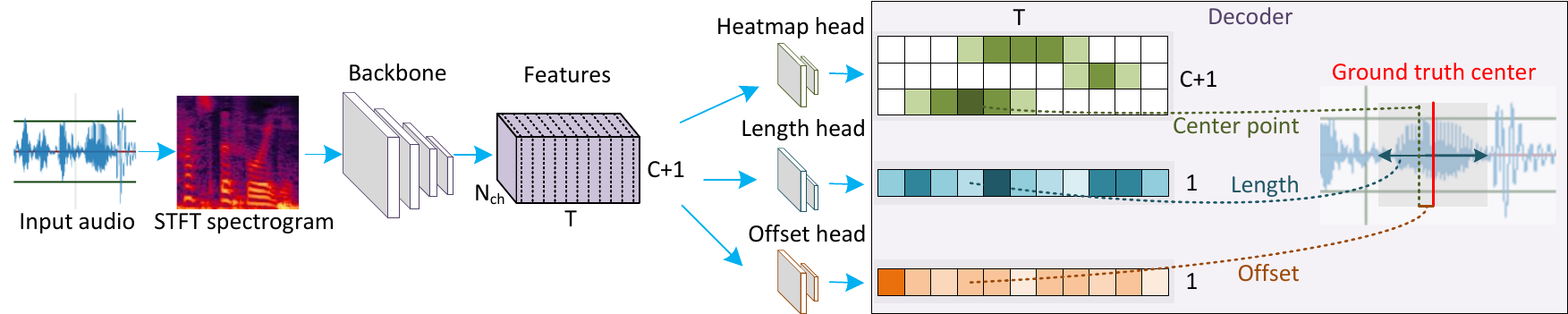}
  \caption{Overview of the proposed anchor-free keywords spotting (AF-KWS) method.}
  \label{fig:overview}
\end{figure*}

Figure~\ref{fig:overview} provides an overview of the proposed method AF-KWS. This end-to-end framework starts with computing Short-Time Fourier Transform (STFT) spectrogram for the input audio. It then employs a trainable backbone to extract the feature map $F \in \mathbb{R}^{T \times (C+1) \times N_{ch}}$, where $T$ is the temporal resolution and $N_{ch}$ is the number of channels. Inspired by the anchor-free detection method in computer vision \cite{other-centernet}, we employ three trainable prediction heads to regress a keyword heatmap $ \hat Y \in {[0,1]^{T \times (C+1)}} $, a keyword length vector $ \hat L \in {\mathbb{R}^T} $, and an offset vector $\hat O \in {\mathbb{R}^T}$, respectively. Offsets are used to correct word positions to compensate for the loss of precision in limited temporal resolution {caused by down-sampling and framing in frequency domain}.

A large $ {{\hat Y}_{t,c}}$ suggests a high probability that a keyword $k_c$ is centered at time index $t$. When the value of $c$ is from 1 to $C$, the detected keyword belongs to the predefined keywords set $K$. However, when $c = C+1$, the detected keyword belongs to an auxiliary class ``unknown" (word). Introducing an auxiliary class presents a different design from what is commonly adopted in visual object detection, where undefined objects and non-objects are all treated as background. We believe that such an auxiliary class which differentiates other words from non-speech background is necessary in CSKWS. This is validated in the ablation studies in Section 4. 

\subsection{Training}

The proposed AF-KWS network is trained in a supervised manner. Thanks to the off-the-shelf 
forced alignment tool\cite{other-mfa}, we are able to label a speech with the locations of each spoken word in an automatic way. 
Let $W = \{w_1, w_2, ..., w_N\}$ be the set of aligned words in an input speech utterance, and let $cls(w_i)$, $loc_{pc}(w_i)$, and $len(w_i)$ be the class label, center location, and length of word $w_i$ ($i=1, ..., N$), respectively. The value of $cls(\cdot)$ is between 1 and $C+1$, and the values of $loc_{pc}(\cdot)$ a real number not exceeding $T$. We further split the precise location of $w_i$ into an integer part $loc(w_i) = \lfloor loc_{pc}(w_i) \rfloor$ and an offset $ofs(w_i) = loc_{pc}(w_i) - loc(w_i)$. For training, we shall convert these automatic alignment results into the ground truth of network output, denoted by $Y$, $L$, and $O$. We shall also define the loss functions for the three prediction tasks.  

The ground truth keyword heatmap $Y$ is a smoothed heatmap. For each word $w_i \in W$, we first set $Y_{loc(w_i), cls(w_i)}$ to 1 and then use a Gaussian kernel to spread the heat to its neighborhood along the temporal axis. The standard deviation $\sigma_{w_i}$ of the Gaussian kernel is proportional to the length of the word, or $ {\sigma_{w_i}} = len(w_i) \times \gamma$, where $ \gamma $ is a size adaptive factor. We set $\gamma=0.125$ based on ablation studies. 

For keypoint heatmap prediction, we use a penalty-reduced pixel-wise logistic regression with focal loss \cite{other-focalloss} 
\begin{equation}
{L_h} = - \frac{{1}}{N}\left\{ {\begin{array}{*{20}{l}}
  {{\text{   }}\sum\limits_{t,c} {{{\left( {1 - {{\hat Y}_{t,c}}} \right)}^\alpha }\log \left( {{{\hat Y}_{t,c}}} \right)} {\text{ if }}{Y_{t,c}} = 1} \\ 
  {\begin{array}{*{20}{l}}
  {\sum\limits_{t,c} {{{\left( {1 - {Y_{t,c}}} \right)}^\beta }{{\left( {{{\hat Y}_{t,c}}} \right)}^\alpha }} } \\ 
  {\sum\limits_{t,c} {\log \left( {1 - {{\hat Y}_{_{t,c}}}} \right)} } 
\end{array}{\,\quad\text{    otherwise}}} 
\end{array}} \right.
\end{equation}
where $ \alpha $ and $ \beta $ are hyper-parameters of the focal loss, and $N$ is the number of keywords in the input audio. 

The groundtruth for keyword length is a $T$-dimensional vector $L$, where $L_t$ is the length of the keyword centered at time index $t$. Note that the temporal resolution $T$ is selected such that the center locations of two adjacent words never appear at the same time index. For each of the $N$ aligned words, we set $L_{loc(w_i)}=len(w_i)$. All the unassigned elements are initialized to zero. 
We use L1 loss for length prediction and it is computed by:
\begin{equation}
{L_{len}} = \frac{1}{N}\sum\limits_{i = 1}^N {\left| {{{\hat L}_{t_i}} - {L_{t_i}}} \right|}
\end{equation}
The offset prediction results provides an adjustment to the center locations of detected keywords. Same as the keyword length prediction, the offset ground truth is a $T$-dimensional vector, denoted by $O \in \mathbb{R}^T$. We set $O_{loc(w_i)}=ofs(w_i)$ for each word $w_i$ and the rest elements are initialized to zero. During training, we use L1 loss for offset prediction:
\begin{equation}
{L_{offset}} = \frac{1}{N}\sum\limits_{i = 1}^N {\left| {\hat O}_{loc(w_i)} - O_{loc(w_i)} \right|}
\end{equation}The overall training loss is:
\begin{equation}
{L_{cskws}} = {L_h} + {\lambda _{len}}{L_{len}} + {\lambda _{offset}}{L_{offset}}
\end{equation}
where $ {\lambda _{len}} $ and $ {\lambda _{offset}} $ is a loss weight factor for length loss and offset loss. We follow previous practice \cite{other-centernet} to set $ {\lambda _{len}} = 0.1 $ and $ {\lambda _{offset}} = 1 $. 


\subsection{Inference process}
During inference, we obtain $\hat{Y}$, $\hat{L}$, and $\hat{O}$ for each audio sample. We need to convert them into a set of detected keywords $\hat{W} = \{\hat{w}_1,...,\hat{w}_M\}$ and associated properties including $cls(\hat{w}_i)$, $loc_{pc}(\hat{w}_i)$, and $len(\hat{w}_i)$. 

With the predicted keywords heatmap $\hat{Y}$, we examine the vector $\hat{Y}_c$ for each keyword class $c$. We identify the peak locations whose response score is greater than its neighbors along the temporal axis. After examine all the keyword classes, we select the top $M$ locations by their response scores. If $\hat{Y}_{t,c}$ is selected, a new keyword $w$ with $cls(w)=c$, $len(w)=\hat{L}_t$, and $loc_{pc}(w)=t+\hat{O}_t$ is added to set $\hat{W}$. Finally, we can obtain a bounding box for this keyword as $\left[ loc_{pc}(w) - len(w)/2, loc_{pc}(w) + len(w)/2 \right]$. $\hat{Y}_{t,c}$ is directly used as the confidence score of this detection result. 
We do not perform non-maximum suppression (NMS) on the output.

\subsection{Implementation details}

The input audio is 16,000 Hz mono audio, and the input length for the model is $r$=5.11s. We first extract STFT features with $ hop\_length = 160 $, $ win\_length = 400 $ and $ filter\_length = 510 $, and then down-sampling them to temporal resolution $T=128$. 
We use up-convolutional residual networks (ResNet) \cite{other-resnet-pose} Res34 as our backbone model. 
The maximum number of detected keywords is set to $ M = 30 $, translating to a speech rate of about 350 words per minute (wpm), which is much faster than a normal speed of 160 to 180 wpm as specified by BBC \cite{other-bbc-speechrate}. In the ablation study, we compare our method with a setting without the auxiliary unknown class. In that case, $M$ is set to 3 or 2, depending on the datasets, to avoid too many false alarms. 
We apply several types of data augmentation, with the same probability of 0.2, to the training data, including additive noise, reverberation, and pitch shifting. 

\section{Dataset}
The availability and quality of benchmark datasets are of great importance to the development of a research topic. There are few dedicated datasets for the CSKWS task. One related dataset is the hybrid real-synthetic football keyword dataset FWD \cite{ckws-football}. An earlier work \cite{ckws-rnn2014} tried to build a dataset by selecting the six most frequently occurring words in TIMIT \cite{data-timit}. However, TIMIT is a small dataset with only 5.4 hours speech. 
In our work, we follow a similar practice as \cite{ckws-rnn2014} to build a much larger benchmark dataset from LibriSpeech. 
We also define and collect a new meeting keywords dataset named CMAK to facilitate meeting-related downstream tasks. 

\begin{table}[t!]
  \caption{Statistics of the LibriTop-20 dataset}
  \label{tab:data-libritop}
  \centering
  \begin{tabular}{lrrr}
    \toprule
    \multicolumn{1}{c}{\textbf{Split}} & 
    \multicolumn{1}{c}{\textbf{Samples}} & 
    \multicolumn{1}{c}{\textbf{Words}} & 
    \multicolumn{1}{c}{\textbf{Keywords}} \\
    \midrule
   Train& 	 251,229& 		  9,521,509&			520,701  \\
   Dev& 		2,953& 		   80,210& 			5,337\\
   Test&			 2,684& 			74,952&			4,602   \\
   Total&			 256,866& 			9,676,671&			530,640   \\
    \bottomrule
  \end{tabular}
\end{table}

\subsection{LibriTop dataset}

The LibriTop dataset is derived from LibriSpeech \cite{data-librispeech}, which is a continuous speech corpus of approximately 1000 hours of 16kHz read English speech. As LibriSpeech is built for the ASR task, all samples in it have been carefully segmented and aligned. We first calculate the word frequencies in LibriSpeech and then try to pick the 20 most frequently used words. However, due to the nature of KWS, it usually cannot be applied to single-syllable words. Therefore, the resulting LibriTop-20 dataset contains 20 top-frequency words with at least two syllables. These words are: Very, Into, Little, About, Only, Upon, Any, Before, Other, Over, After, Never, Our, Mister, Again, Himself, Away, Even, Without, Every.

There are a total of 256k utterances and 530k occurrences of keywords in LibriTop-20. We keep the same data split as LibriSpeech. The detailed statistics are listed in Table~\ref{tab:data-libritop}.

\begin{table}[tb!]
  \caption{Statistics of the CMAK dataset}
  \label{tab:data-cmak}
  \centering
  \begin{tabular}{lrr}
    \toprule
    \multicolumn{1}{c}{\textbf{Source}} & 
    \multicolumn{1}{c}{\textbf{Speakers}} & 
    \multicolumn{1}{c}{\textbf{Utterances}} \\
    \midrule
   Recorded& 	 32& 		  15,092  \\
   Synthetic& 		136& 		   334,847\\
   Total&			 168& 			349,939   \\
    \bottomrule
  \end{tabular}
\end{table}

\setlength{\tabcolsep}{5pt}
\begin{table*}[htbp]
  \caption{Performance comparison between the proposed AF-KWS and the reference schemes on LibriTop-20. AP@K means the AP at an IoU threshold of K\%. FRR@K means the FRR when the FAs per hour is K. }
  \label{tab:exp_cmp}
  \centering
  \begin{tabular}{lccccccccc}
    \toprule
     \multicolumn{1}{c}{\textbf{Model}} & 
    \multicolumn{1}{c}{\textbf{AP@5$\uparrow$}} &
    \multicolumn{1}{c}{\textbf{AP@75$\uparrow$}} &
    \multicolumn{1}{c}{\textbf{mAP$\uparrow$}} &
    \multicolumn{1}{c}{\textbf{FRR@5$\downarrow$}} &
    \multicolumn{1}{c}{\textbf{FRR@15$\downarrow$}} &
    \multicolumn{1}{c}{\textbf{FRR@25$\downarrow$}} &
      \multicolumn{1}{c}{\textbf{Classification Accuracy$\uparrow$}} &
    \multicolumn{1}{c}{\textbf{RTF$\downarrow$}} \\
    \midrule
    DSTC-ResNet&		0.748&		0.058&		0.398&		0.647&	0.519&		0.402&	    0.961&		0.018\\
    MHAtt-RNN&			0.795&		0.076&			0.426&			0.530&			0.418&				0.374&	    0.978&		0.057\\
    AF-KWS (ours)&		0.952&			0.886&			0.860& 		0.140&				0.074&		0.049&      N/A&		0.031\\
    \bottomrule
  \end{tabular}
\end{table*}

\subsection{CMAK dataset}

CMAK dataset is collected for speech-based meeting structure analysis. We defined a total of 24 keywords, relating to the start, the end, or some specific sections of a meeting. They are:
\begin{itemize}[leftmargin=*]
\item Start: Begin, Start, Agenda, Outline, Today, Introduce, Talk about
\item End: Summarize, Conclude, Action item, To do, Follow up, Next step, Share, Plan
\item Specific sections: Timeline, Milestone, Question, Comments, Additions, Highlight, Important, Crucial, Attention
\end{itemize}

CMAK is a hybrid real-synthetic dataset. Different from FKD which contains hybrid utterances, each utterance in CMAK is either real or synthetic. To collect the real data, we select more than 15 sentences from the Oxford English dictionary \cite{other-oxxford} for each keyword, and ask 32 speakers to read the scripts and record the speech. To collect the synthetic data, we randomly generate a script with $ 10 \sim 15 $ words containing one keyword at a time, and employ ESPNet \cite{other-espent,other-espnet-tts,other-espnet-st,other-espnet-se,other-espnet-slu} to synthesize the speech. We use 136 different speaker embeddings to generate $ 100 $ sentences for each keyword. The detailed statistics for the dataset are listed in Table~\ref{tab:data-cmak}.

We carry out most of the experiments in this paper using a subset of CMAK, called CMAK-7. It contains the seven keywords relating to the start of a meeting. A model trained on this subset can be applied to the first few minutes of a recorded meeting to enable the \textit{Skip greetings} feature. Further, CMAK-7-s is a subset of CMAK-7 that only contains the synthetic speech.

\section{Experiments}

We evaluate the proposed AF-KWS method on LibriTop-20 and CMAK-7 datasets. LibriTop-20 is divided in the same way as LibriSpeech, and CMAK-7 is divided into $ 80\% $  training set and $ 20\% $ testing set. For different versions of a method, the input length is fixed. We apply repeated padding for shorter input and random crop for longer input during training. {We train the model with a batch size of 64 and a learning rate of 0.00125.}

\subsection{Evaluation metrics}

In trigger words detection, false rejection rate (FRR) and false accepts per hour (FAs) are commonly used as the evaluation metrics \cite{wake-google-contextual, wake-snip-efficient, wake-xiaoaitongxue, wake-amazon}. In speech command recognition, accuracy is commonly used for evaluation \cite{trim-bcresnet, trim-sbc,trim-tcn}. CSKWS can borrow the evaluation metrics from these related tasks. However, there are more keywords in CSKWS and they appear more frequently than in trigger words detection. Besides, CSKWS is essentially a detection task instead of classification task, so we propose to evaluate its solutions with object detection metrics known as 1D AP(average precision) and mAP
(mean average precision). 
1D IoU (Intersection over Union) is computed for each detection result with respect to the groundtruth, and mAP is computed in an IoU range of $ (0.05, 0.95) $ by step 0.05. This is a slightly different setting from the mAP used in visual object detection \cite{other-map}, because there is a higher tolerance of positional accuracy in CSKWS.

\setlength{\tabcolsep}{4pt}
\begin{table}[tb!]
  \caption{Ablation study on LibriTop-20}
  \label{tab:exp_ab}
  \centering  
    \begin{tabular}{lcccc}
    \toprule
     \multicolumn{1}{c}{\textbf{Model}} & 
    \multicolumn{1}{c}{\textbf{AP@5}} &
    \multicolumn{1}{c}{\textbf{AP@75}} &
    \multicolumn{1}{c}{\textbf{FRR@5}} &
    \multicolumn{1}{c}{\textbf{FRR@25}} \\
    \midrule
    AF-KWS-cls &			0.876&			0.062&				0.570&					0.265      \\
    w/o unknown&			0.867&			0.800&				0.234&					0.117    \\
    AF-KWS (ours)&		0.952&			0.886&				0.140&					0.049   	\\
    \bottomrule
  \end{tabular}
\end{table}

\subsection{Performance comparison}

CSKWS is a less studied task than speech command recognition, or wake-up words detection. There are few reference schemes and none of them have publicly available codes. Therefore, we choose to adapt the SOTA speech commands recognition methods for CSKWS task and use them as reference schemes. We pick two Google implemented models \cite{trim-google-streaming}, MHAtt-RNN \cite{trim-google-streaming} and DSTC-ResNet based on MatchboxNet \cite{trim-matchbox}, which achieve top results on Google speech command dataset. 

We insert a sliding window region proposal module before these speech commands recognition methods, which are essentially classifiers. The sliding window strategy has two parameters, namely the window size $ {L_{in}} $ and the sliding step $ {L_{step}} $.  
To make sure that each keyword is included in at least one window, we apply the constraints that $ {L_{in}} > MA{X_x} $ and $ {L_{step}} < {L_{in}} - MA{X_x} $ , where $ MA{X_x} $ is the maximum length of the keywords. We set $L_{in}=0.5$ for LibriTop-20. There is a trade-off between detection performance and inference speed, so we grid search the best step size from 0.1s to 0.4s with 0.1s interval. 
We re-train the adapted KWS methods on our dataset. 

Table~\ref{tab:exp_cmp} shows the comparison between AF-KWS and the two reference schemes. We list the results on both detection-oriented metrics and conventional KWS metrics. 
Both DSTC-ResNet and MHAtt-RNN can achieve a high classification accuracy on trimmed input, showing that our re-training is successful. 
However, they give poor AP numbers, because sliding window region proposal cannot provide high temporal resolution, or the inference cost will be too high. 
The results show that our AF-KWS model outperforms the two methods on AP and FRR with almost same inference speed \footnote{The real time factor (RTF) is tested on our local CPU machine with Intel Xeon CPU E5-2690 v4 @2.6GHz.}.

\subsection{Ablation experiments}
Our AF-KWS detector outperforms two reference schemes adapted from classifiers. To verify that the gain is not due to a stronger backbone, we carry out an ablation study which replaces the three detection heads in AF-KWS with a classification head. Accordingly, the input length is adjusted to 0.5s and a sliding window inference strategy is adopted. Table~\ref{tab:exp_ab} shows that this AF-KWS-cls model does not perform well, confirming our claim that CSKWS should be treated as a detection task. 

AF-KWS is inspired by anchor-free visual object detectors, but we make one key design choice that differentiates it from normal detectors. AF-KWS defines an auxiliary ``unknown" class in addition to the predefined keyword classes. The reason is that if only the defined keywords are treated as foreground, the background will include other words, non-speech sound, and silence, which is too diverse to handle. To verify this, we implement a ``w/o unknown" version of AF-KWS. As Table~\ref{tab:exp_ab} shows, its performance is significantly inferior to our method.

Finally, we study the impact of dataset using different versions of CMAK-7. CMAK is a hybrid dataset with both real recordings and synthetic speech. Due to privacy concerns, we may only able to release the synthetic part of CMAK. In Table~\ref{tab:exp_ab_cmak}, the results of the first two rows show that our proposed model achieves similar results on hybrid and synthetic datasets, given that the test set has the same data distribution as the training set. The last row shows the results when we use additional LibriSpeech data for training. The performance is significantly boosted on the same test set as used in the second row. 
\setlength{\tabcolsep}{2pt}
\begin{table}[t!]
  \caption{Performance on CMAK-7 with different datasets}
  \label{tab:exp_ab_cmak}
  \centering  
     \begin{tabular}{lccccc}
    \toprule
     \multicolumn{1}{c}{\textbf{Dataset}} & 
     \multicolumn{1}{c}{\textbf{Extra Data}} & 
    \multicolumn{1}{c}{\textbf{AP@5}} &
    \multicolumn{1}{c}{\textbf{AP@75}} &
    \multicolumn{1}{c}{\textbf{FRR@5}} &
    \multicolumn{1}{c}{\textbf{FRR@25}} \\
    \midrule
    Synthetic& N &		0.859&			0.697&				0.360&				   0.137\\
    Hybrid& N &		0.874&			0.633&				0.356&				   0.159\\
    Hybrid&	Y &	0.916&			0.731&				0.238&				   0.126\\
\bottomrule
  \end{tabular}
  
\end{table}

\section{Conclusion}

In this paper, we have discussed the continuous speech keyword spotting (CSKWS) task which finds many important applications in the post-pandemic era. We have introduced the LibriTop-20 dataset derived from LibriSpeech and have collected a new continuous meeting analysis keywords dataset named CMAK. We have proposed a novel anchor-free keywords detection method which is tailored for the CSKWS task. The proposed AF-KWS method achieves better performance than the adapted speech command recognition methods, which verifies our claim that CSKWS should be treated as a detection problem instead of a classification problem. AF-KWS provides a reasonable baseline for future research on this interesting CSKWS problem. 

\bibliographystyle{IEEEtran}
\bibliography{mybib}
\end{document}